\documentclass[aps,prd,showpacs,showkeys,twocolumn]{revtex4}
\usepackage{epsfig,amsmath,amsfonts,amssymb}
\begin{document}
\title{\bf Skyrme black holes in the isolated horizons formalism}
\author{Alex B. Nielsen}
\email{abn16@student.canterbury.ac.nz}
\affiliation{Department of Physics and Astronomy, University of
Canterbury,
Private Bag 4800, Christchurch, New Zealand}
\begin{abstract}
We study static, spherically symmetric, Skyrme black holes in the
context of the assumption that they can be viewed as bound states
between ordinary bare black holes and solitons. This assumption
and results stemming from the isolated horizons formalism lead to
several conjectures about the static black hole solutions. These
conjectures are tested against the Skyrme black hole solutions. It
is shown that, while there is in general good agreement with the
conjectures, a crucial aspect seems to violate one of the
conjectures.

\centerline{gr-qc/0603127}

\medskip

\centerline{3 April 2006; \LaTeX-ed \today }
\end{abstract}
\pacs{04.70.-s, 04.70.Bw, 04.70.Dy, 12.39.Dc} \keywords{Black
holes, isolated horizons, skyrmions;}
\maketitle
\newcommand{\beq}{\begin{equation}}
\newcommand{\eeq}{\end{equation}}
\newcommand{\bea}{\begin{eqnarray}}
\newcommand{\eea}{\end{eqnarray}}
\def\d{{\mathrm{d}}}
\def\Tr{{\mathrm{Tr}}}
\section{Introduction}

The isolated horizons formalism \cite{Ashtekar:2000sz} has been
used to formulate several conjectures regarding static black holes
\cite{Ashtekar:2000nx,Corichi:2000dm}. In this article we study
the static, spherically symmetric Einstein-Skyrme black hole
solutions first found in
\cite{Luckock:1986em,Bizon:1992gb,Droz:1991cx}. The
Einstein-Skyrme solutions have a number of features generic to
certain models involving non-abelian gauge fields
\cite{Volkov:1998cc} that complement features of models previously
considered in the context of the isolated horizons conjectures
\cite{Kleihaus:2000kv,Breton:2003tk,Corichi:2005pa,Ibadov:2005rb}.
Like a large class of models admitting solitons in flat space,
Skyrme black holes have an upper limit on the radius of the black
hole. This involves the merging of two different branches of
solutions at some finite horizon radius. In addition they admit
static solutions that are not spherically symmetric
\cite{Sawado:2004yq}, thus extending the class of static
solutions. Perhaps most importantly Skyrme black holes are of
particular interest in the context of the no-hair theorem since
they have been shown to be linearly stable
\cite{Bizon:1992gb,Heusler:1992av,Heusler:1993ci} and thus
demonstrate that exceptions to the no-hair theorem are not
necessarily unstable. However, it has yet to be resolved whether
Skyrme black holes are non-linearly stable and it is possible that
the conjectures stemming from the isolated horizon formalism could
shed some light on this unresolved issue. One of the advantages of
the isolated horizons formalism is that it characterizes black
hole solutions in terms of horizon values of their parameters
rather than their values at infinity. Thus the formalism provides
a well-suited framework in which to investigate so-called ``hairy"
black holes, where solutions that are indistinguishable at
infinity have different parameter values at the horizon.

This paper is organized as follows: Section \ref{sec:skyrmeeqns}
reviews the basics of the Skyrme model and the black hole
solutions found previously
\cite{Luckock:1986em,Bizon:1992gb,Droz:1991cx}. Section
\ref{sec:ihconjecs} reviews the essential elements of the isolated
horizon formalism and the conjectures for static black holes that
it gives rise to. Section \ref{sec:results} presents our results
and a discussion of their relevance to the conjectures. Our metric
convention is $(-,+,+,+)$ and we use units where $c=1$, although
since we often want to compute masses in terms of spatial
integrals we keep all factors of $G$. Otherwise our conventions
are as appear in Wald \cite{Wald:book}.

\section{Static Skyrme Equations}\label{sec:skyrmeeqns}
The lagrangian density for the Skyrme model is
\beq \label{lskyrme}
{\cal{L}}=-\frac{1}{4}f^{2}\Tr(A_{a}A^{a})+\frac{1}{32g^{2}}\Tr(F_{ab}F^{ab})
\eeq
where $A_{a} = U^{\dagger}\nabla_{a}U = U^{\dagger}\partial_{a}U$
and $F_{ab} = A_{a}A_{b}-A_{b}A_{a}$, and $f$ and $g$ are coupling
constants. The static, spherically symmetric ansatz (hedgehog
ansatz) for the $SU(2)$ valued Skyrme field is
\beq U(r) = \cos \chi(r) + i\sin \chi (r) \mathbf{\tau}.\hat{x} =
e^{i\chi(r)\mathbf{\tau}.\hat{x}}\eeq
where $\mathbf{\tau}.\hat{x} = \tau_{r}$, $\hat{x}^{a} =
(\sin\theta\cos\phi, \sin\theta\sin\phi,\cos\theta)$ and
$\tau_{a}$ are the usual Cartesian Pauli matrices. Thus
\bea \label{Askyrme} -iA & \equiv & -iA_{\mu}\d x^{\mu} =
\tau_{r}\chi'\d r \nonumber \\ & + &
\left[\cos\chi\sin\chi(\partial_{\theta}\tau_{r}) -
i\sin^{2}\chi(\tau_{r}\partial_{\theta}\tau_{r})\right]\d\theta
\nonumber \\ & & + \left[\cos\chi\sin\chi(\partial_{\phi}\tau_{r})
- i\sin^{2}\chi(\tau_{r}\partial_{\phi}\tau_{r})\right]\d\phi
.\eea
The Pauli matrices
$\tau_{r},\tau_{\theta}=\partial_{\theta}\tau_{r},\tau_{\phi} =
\frac{1}{\sin\theta}\partial_{\phi}\tau_{r}$ satisfy the usual
commutation relations $[\tau_{r},\tau_{\theta}] =2i\tau_{\phi}$
(and cyclical permutations) and
$\tau_{r}^{2}=\tau_{\theta}^{2}=\tau_{\phi}^{2}=1$. The
energy-momentum tensor for the Skyrme lagrangian density above is
\bea \label{Tskyrme} T_{ab} & = & -\frac{1}{2}f^{2}\Tr(A_{a}A_{b}
- \frac{1}{2}g_{ab}A_{a}A^{b}) \nonumber \\ & & +
\frac{1}{8g^{2}}\Tr(F_{ac}F_{bd}g^{cd} -
\frac{1}{4}g_{ab}F_{cd}F^{cd}) .\eea
Note that the trace of the energy-momentum tensor is in general
non-zero. The fact that the Skyrme model has its own length scale
permits the existence of flat-space solitons. It has been
conjectured that it is this length scale that is responsible for
the upper bound on the size of the black hole within the soliton
as has also been observed in other theories that admit flat-space
solitons (see \cite{Hartmann:2001ic,Volkov:1998cc} and references
therein). This is not the case for the Einstein-Yang-Mills black
holes considered in \cite{Corichi:2000dm}, since Yang-Mills theory
does not admit flat space solitons and the trace of its
energy-momentum tensor is always zero. However, in the
Einstein-Yang-Mills-Higgs model considered in
\cite{Corichi:2005pa}, the trace of the energy-momentum tensor is
generically non-zero and black holes exist of arbitrary size.

The baryon current is given by
\beq b^{a} = \frac{1}{24\pi^{2}}\epsilon^{abcd}\Tr
(U^{\dagger}\nabla_{b}UU^{\dagger}\nabla_{c}UU^{\dagger}\nabla_{d}U)
.\eeq
The topological charge on a given hypersurface $\Sigma$ is then
\bea B & = & \int_{\Sigma}B^{a}n_{a}\d S \nonumber \\ & = &
\int_{\Sigma} \d^{3}x\sqrt{-g}B^{0} \eea
where $\d S$ is the volume form for $\Sigma$ and $n_{a}$ is the
unit normal to the hypersurface. Thus, the spherically symmetric
hedgehog ansatz above, for the regular soliton, gives
\beq \label{Bdefn} B = \frac{1}{\pi}\left[\chi(r)-\frac{1}{2}\sin
2\chi(r)\right]^{0}_{\infty} .\eeq
Consequently for a regular Skyrme soliton (Skyrmion) the baryon
number will be an integer. In the black hole case it has been
argued that the integration should only run from the horizon
$r_{\triangle}$ to $\infty$ and hence the black hole solutions do
not have integer baryon number \cite{Bizon:1992gb}. In addition,
for $B=n>1$ the solutions will not be the lowest energy
configurations and thus will probably not be stable to decay into
$n$ well isolated $B=1$ skyrmions.  In particular, the $B=2$
lowest energy configuration is known to be axially symmetric, the
black hole versions of which have been given in
\cite{Sawado:2004yq}.\bigskip

\noindent The ansatz for a spherically symmetric metric can be
written as
\bea \label{massmetric} \d s^{2} & = &
-A^{2}(r)\left(1-\frac{2Gm(r)}{r}\right)\d t^{2} \nonumber \\ & &
+
 \left(1-\frac{2Gm(r)}{r}\right)^{-1}\d r^{2}+r^{2}d\Omega^{2}
.\eea
The $G^{t}_{\hspace{0.05cm}t}$ and $G^{r}_{\hspace{0.05cm}r}$
components of the Einstein equations for matter given by
(\ref{Askyrme}) and (\ref{Tskyrme}) are;
\bea \frac{2m'}{r^{2}} & = & 4\pi  \bigg[
f^{2}\left(\chi'^{2}(1-\frac{2Gm}{r})+\frac{2}{r^{2}}\sin^{2}\chi\right)
\nonumber
\\ & & +
\frac{1}{g^{2}}\left(\frac{2}{r^{2}}\chi'^{2}(1-\frac{2Gm}{r})\sin^{2}\chi
+ \frac{\sin^{4}\chi}{r^{4}}\right)\bigg] \eea
and
\beq \frac{4A'm-2A'r}{r^{2}A}+\frac{2m'}{r^{2}} = \nonumber \eeq
\bea 4\pi \bigg[
f^{2}\left(-\chi'^{2}(1-\frac{2Gm}{r})+\frac{2}{r^{2}}\sin^{2}\chi\right)
\nonumber \\  +
\frac{1}{g^{2}}\left(-\frac{2}{r^{2}}\chi'^{2}(1-\frac{2Gm}{r})\sin^{2}\chi
+ \frac{\sin^{4}\chi}{r^{4}}\right)\bigg] \eea
The $G^{r}_{\hspace{0.05cm}r}$ equation can be more simply written
as
\beq \label{Adash} \frac{A'}{A}\frac{1}{r} = 4\pi
G\left[f^{2}(\chi')^{2}+\frac{2}{g^{2}r^{2}}(\chi')^{2}\sin^{2}\chi\right]
.\eeq
Note that $A'$ is always positive if $A$ is initially positive.
These equations of motion should be supplemented by the boundary
conditions for asymptotically flat black hole solutions;
\bea \label{boundarycond} m(r_{\triangle}) & = & \frac{r_{\triangle}}{2G} \nonumber \\
A(r_{\infty}) & = & 1 \nonumber \\
\frac{2Gm(r)}{r}|_{r=\infty} & = & 0 \nonumber \\
\chi(r_{\infty}) & = & 0 .\eea
The shooting parameter for the black hole to be used in the
numerical integration is $\chi(r_{\triangle})$. As is shown in
\cite{Bizon:1992gb} this set of equations will have solutions for
all horizon values $r_{\triangle}$ up to some maximum
$r_{\triangle ,max}$ and this $r_{\triangle ,max}$ will depend on
the dimensionless coupling constant $\alpha = 4\pi Gf^{2}$. A
smaller $\alpha$ will give larger maximum radii and in the limit
$\alpha \rightarrow 0$ the Skyrme field totally decouples from
gravity and we recover the Schwarzschild solution with no upper
bound on the black hole radius. The maximum radius $r_{\triangle
,max}$ will also depend on the topological baryon number $B$. A
larger baryon number black hole will have a lower maximum radius.
For a given $r_{\triangle}$, less than $r_{\triangle ,max}$, there
will be, in general, several classes of solutions, each class
labelled by a unique baryon number and each class containing a
stable branch and an unstable branch. In the limit of
$r_{\triangle} \rightarrow 0$ the two branches will tend to the
two separate solutions of the regular soliton case
\cite{Bizon:1992gb}.

For the regular Skyrmion case, the boundary condition for
$\chi(r)$ in (\ref{boundarycond}) has the result of making $U(r)$
a mapping from $S^{3}$ into $SU(2)$ characterized by the third
homotopy group of $SU(2)$, $\pi_{3}(SU(2))$. Thus they fall into
topologically distinct equivalence classes labelled by an integer
$B$ given by (\ref{Bdefn}). Solutions with different integer
values of $B$ are therefore topologically distinct and cannot be
deformed into one another. However, in the case of black hole
solutions, the solution is only defined on a domain that excludes
a central ball (the black hole) and thus all the mappings are
topologically trivial. Thus there is no expectation of
conservation of the, now non-integer, baryon number $B$ and it is
possible that the stable Skyrme black hole solution could decay
into a Schwarzschild black hole via quantum tunnelling
\cite{Bizon:1992gb}. It is also possible that a Skyrme black hole
with $B>1$ could decay into several widely separated Skyrmions and
a black hole or indeed a spherically symmetric Skyrme black hole
with $B>1$ could decay into a non-spherically symmetric black hole
\cite{Sawado:2004yq}.

\section{Isolated Horizons Conjectures}\label{sec:ihconjecs}
\subsection{Surface Gravity}
First of all we need to define the surface gravity.  This will in
general depend on a normalization. For a static, spherically
symmetric black hole we can use the fact that the horizon will be
a Killing horizon and thus the surface gravity is just given by
\beq \kappa = \lim_{r \rightarrow r_{H}}\left(
\frac{1}{2}\frac{\partial_{r}g_{tt}}{\sqrt{g_{tt}g_{rr}}}\right)
.\eeq
For the form of the metric given explicitly in terms of the mass
function (\ref{massmetric}) we have
\beq \kappa = \kappa_{s}A(1-2Gm') .\eeq
In order to chose a normalization such that the time-translational
Killing vector becomes unity at infinity the value of $A$ (which
is really just pure gauge since it does not affect the dynamics)
should be set equal to one at infinity.  This fixes the overall
normalization for the surface gravity.
\subsection{Horizon Mass and Conjectures}
The conjectures for static black holes start from the physical
idea that a hairy black hole can be viewed as a bound state of an
ordinary hairless black hole and a soliton of the matter theory,
in this case a Skyrmion \cite{Ashtekar:2000nx,Corichi:2000dm}.
Using the fact that the value of the total Hamiltonian is constant
on any connected component of static solutions, the
Arnowitt-Deser-Misner (ADM) mass of a given black hole solution
should be decomposable into a mass associated with the horizon and
a mass associated with the soliton;
\beq \label{boundmasses}
M_{\mathrm{ADM}}=M_{\mathrm{sol}}+M_{\mathrm{\triangle}} .\eeq
The ADM mass is simply $m(\infty)$. Taking the isolated horizon to
be an internal boundary for the spacetime manifold, the horizon
mass can be chosen to take the form
\beq M_{\triangle} =
\frac{1}{2G}\int_{0}^{r_{\triangle}}\beta(\tilde{r}_{\triangle})
\d \tilde{r}_{\triangle}\eeq
where $\beta = 2r_{\triangle}\kappa = A(1-2Gm')$ and the
integration should be taken over the black hole solutions with
horizon radii up to $r_{\triangle}$ (see \cite{Ashtekar:2000nx}
and references therein). Each branch of solutions can be labelled
by an integer $n$. Since for each value of $B$ there is both a
stable branch and an unstable branch, we can take $n=2B-1$ for the
stable branches and $n=2B$ for the unstable branches. While $B$ is
non-integer for the black holes, this will still be
`approximately' true since the values do not deviate much from the
integer values. Thus for a given horizon radius $r_{\triangle}$ on
a given branch $n$
\bea M_{\mathrm{ADM}}^{(n)} & = & \nonumber
\\ & & \hspace{-2cm} M_{\mathrm{sol}}^{(n)}+\frac{1}{2G}\int_{0}^{r_{\triangle}}\d
\tilde{r}_{\triangle}\left[A^{(n)}\left(1-2G\frac{\d m}{\d
r}^{(n)}\right)\right]_{r=\tilde{r}_{\triangle}}.\eea
Writing the ADM mass as the sum of the two constituent parts and
their binding energy,
\beq
M_{\mathrm{ADM}}^{(n)}=M_{\mathrm{sol}}^{(n)}+M_{\triangle}^{(0)}
+ E_{bind} \eeq
gives the binding energy as
\beq E_{bind} = M_{\triangle}^{(n)} - M_{\triangle}^{(0)} .\eeq
This provides the motivation for the following conjectures
\footnote{While these conjectures were first proposed in the
context of colored, that it is to say Einstein-Yang-Mills, black
holes it is stated in section 4. of \cite{Ashtekar:2000nx} that
they should apply to more general black holes, such as
Einstein-Skyrme, provided care is taken over the definition of the
horizon mass at the crossover point.}. Firstly, since the binding
energy must be negative,
\begin{enumerate}
    \item [{\bf{1.}}] $M_{\triangle}^{(n)}(r_{\triangle}) <
    M_{\triangle}^{(0)}(r_{\triangle})$
    for all $n>0$ and all $r_{\triangle}$
\end{enumerate}
The definition of the horizon mass then implies:
\begin{enumerate}
\item [{\bf{2.}}]$\kappa^{(n)}(r_{\triangle}) <
\kappa^{(0)}(r_{\triangle})$ for all $n>0$ and all
    $r_{\triangle}$.
\end{enumerate}
Since the magnitude of the gravitational binding energy should
increase (i.e. the binding energy should become more negative) as
the mass of either of the two bound objects grows, when we
consider fixing the radius $r_{\triangle}$ of the black hole and
increasing $n$, we get:
\begin{enumerate}
    \item [{\bf{3.}}] For a fixed value of $r_{\triangle}$ the horizon mass
    $M^{n}_{\triangle}$ and the surface gravity $\kappa^{n}$ are
    monotonically decreasing functions of $n$.
\end{enumerate}
When fixing the mass of the soliton and increasing the mass of the
black hole by increasing $r$, we get:
\begin{enumerate}
    \item [{\bf{4.}}] $\beta ^{(n)}(r_{\triangle}) < 1$ for all $n>0$ and all
    $r_{\triangle}$.
\end{enumerate}
Since $M_{ADM}$ is monotonically increasing with $r_{\triangle}$
and $M_{sol}$ is fixed for fixed $n$, by (\ref{boundmasses}):
\begin{enumerate}
    \item [{\bf{5.}}] $M_{\triangle}^{(n)}$ is a monotonically increasing function
    of $r_{\triangle}$, is positive for all values of $n$ and
    vanishes at $r_{\triangle} = 0$.
\end{enumerate}
As mentioned earlier one of the key features that distinguishes
Einstein-Skyrme black holes from their Einstein-Yang-Mills
counterparts is the existence of an upper radius for black hole
solutions.  This maximum radius forms a `crossing point' at which
two different branches of solutions meet, one of which is stable
to linear perturbations and one of which is unstable. Using the
expression for the ADM mass in terms of the soliton mass and
horizon mass and requiring the ADM mass to be uniquely defined
even at the crossing point one obtains \cite{Ashtekar:2000nx}
\bea \label{crossint} M_{sol}^{stab}-M_{sol}^{unstab} & = &
\frac{1}{2G}\int_{0}^{r_{max}}\beta^{stab}(r_{\triangle})\d
r_{\triangle} \nonumber \\ & & -
\frac{1}{2G}\int_{0}^{r_{max}}\beta^{unstab}(r_{\triangle})\d r_{\triangle} \nonumber \\
& = & \frac{1}{2G}\oint\beta(r_{\triangle})\d r_{\triangle} \eea
where the integral should be taken along the closed contour formed
by the two branches of solutions that meet at the crossing point
and the vertical axis between their endpoints at
$r_{\triangle}=0$.
\section{Results}\label{sec:results}
\begin{figure}[ht]
\epsfig{figure=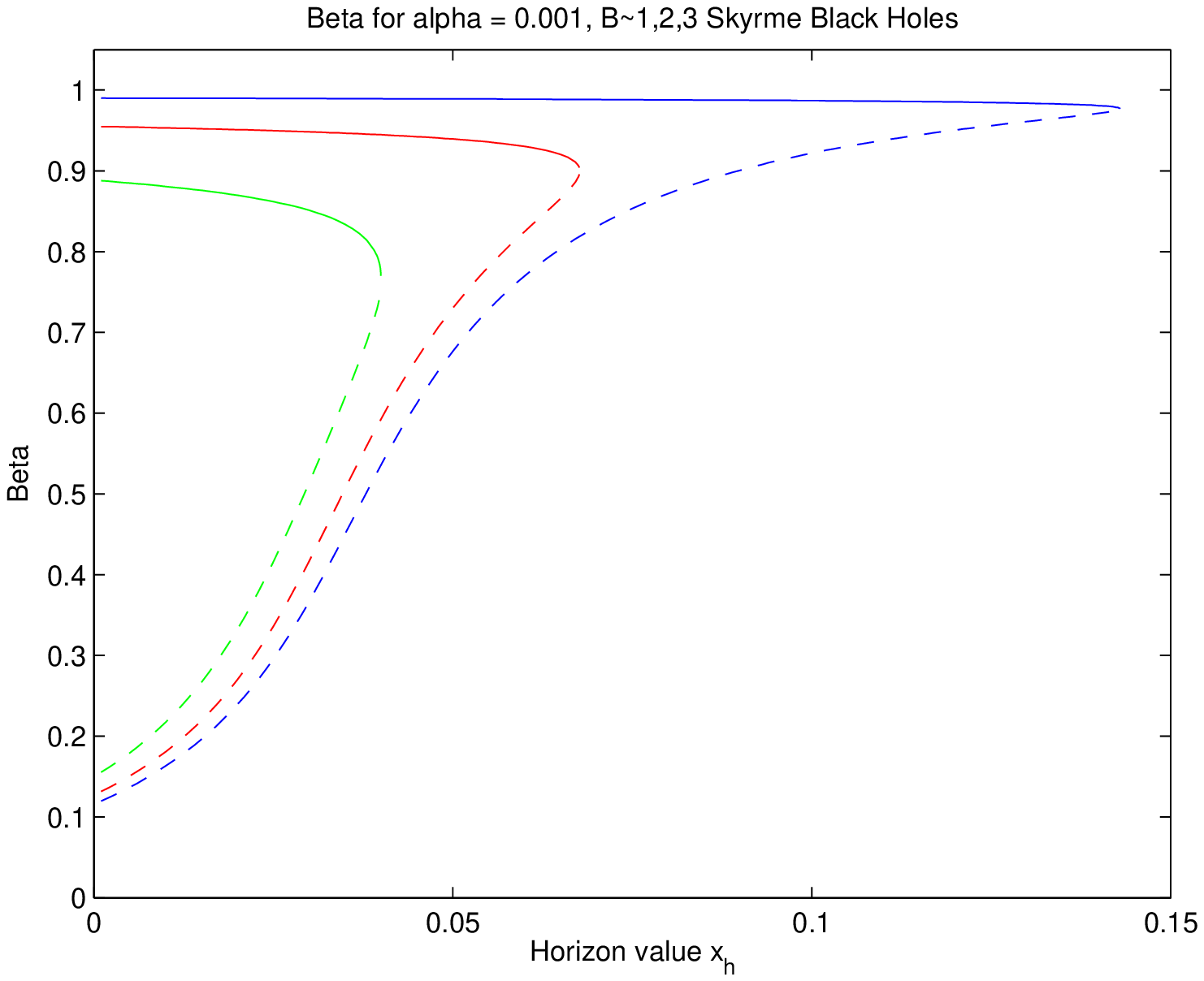,height=8cm,width=7cm}
\caption{$\beta(x_{\triangle})$ for the two branches of the $B\sim
1$ (blue line), $B\sim 2$ (red line) and $B\sim 3$ (green line)
Skyrme Black Holes with $\alpha = 4\pi Gf^{2} = 0.001$. The dotted
lines are the unstable solutions.} \label{k001betas}
\end{figure}
\begin{figure}[hb]
\epsfig{figure=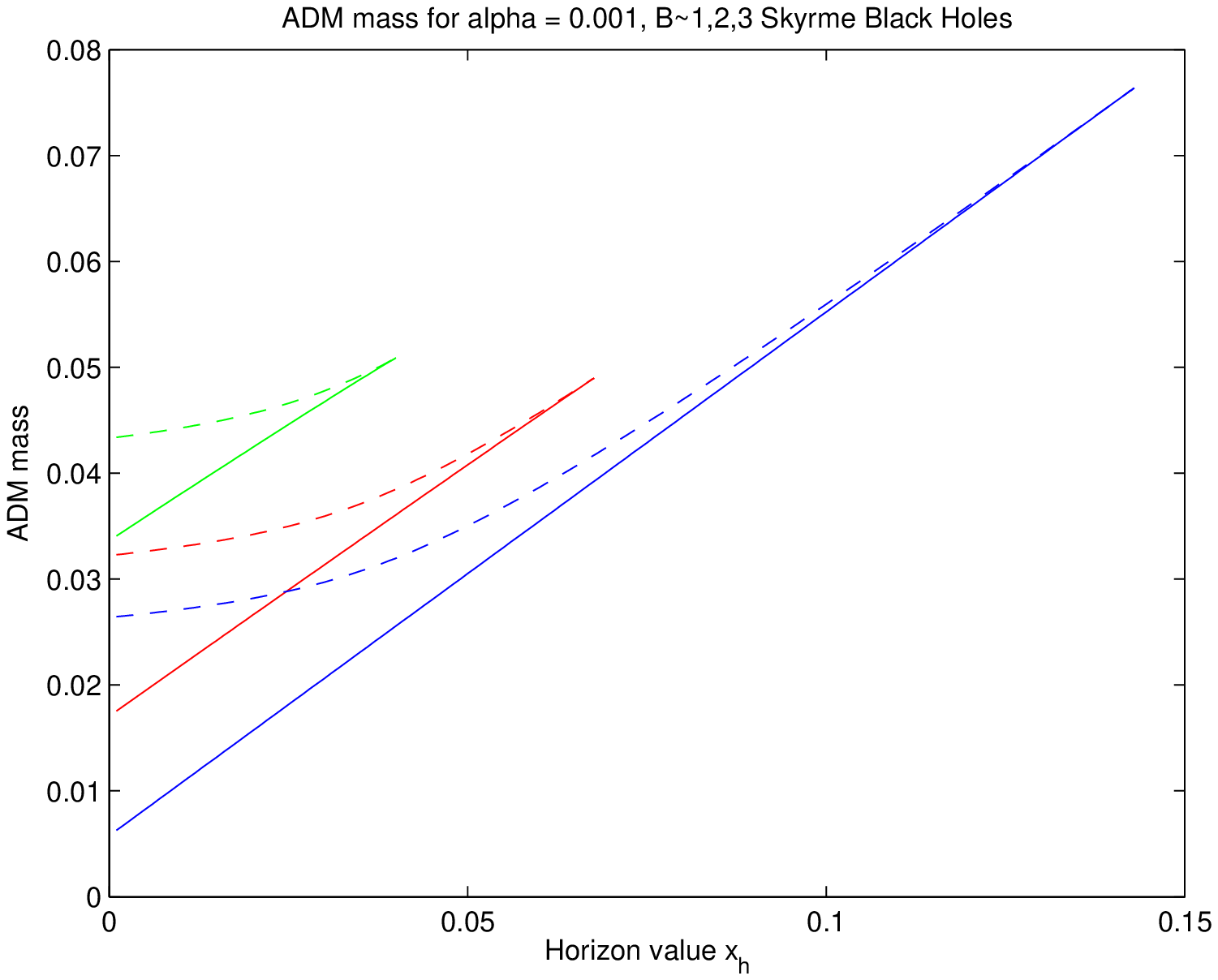,height=8cm,width=7cm} \caption{The ADM
masses for the two branches of the $B\sim 1$ (blue line), $B\sim
2$ (red line) and $B\sim 3$ (green line) Skyrme Black Holes with
$\alpha = 4\pi Gf^{2} = 0.001$. The dotted lines are the unstable
solutions.}\label{k001ADMmasses}
\end{figure}
\begin{figure}[ht]
\epsfig{figure=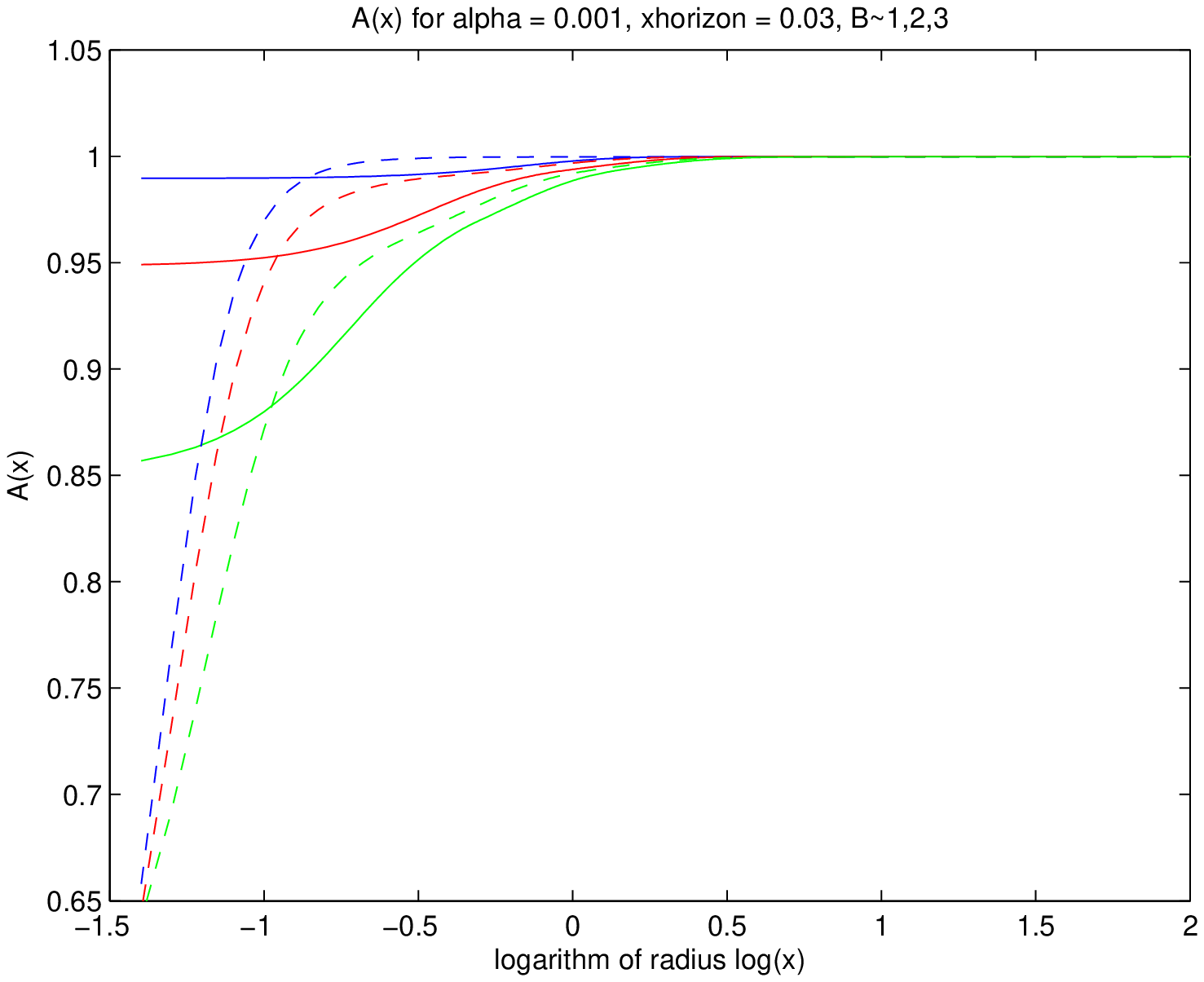,height=8cm,width=7cm} \caption{The
function $A(x)$ versus dimensionless length $x=fgr$ for the two
branches of the $B\sim 1$ (blue line), $B\sim 2$ (red line) and
$B\sim 3$ (green line) Skyrme Black Holes with $\alpha = 4\pi
Gf^{2} = 0.001$ and $x_{\triangle}=0.03$. The dotted lines are the
unstable solutions.}\label{Axh03}
\end{figure}
\begin{figure}[ht]
\epsfig{figure=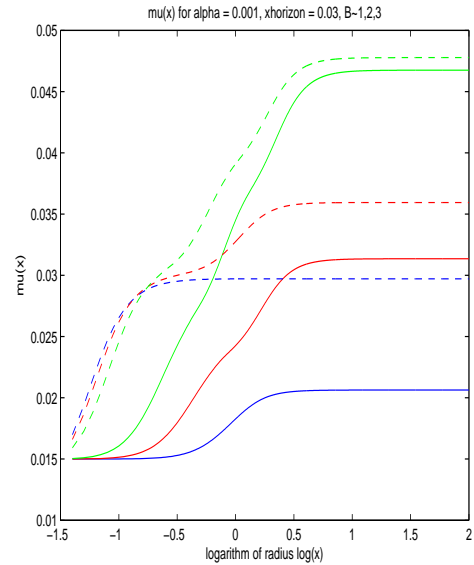,height=8cm,width=7cm} \caption{The
dimensionless mass function $\mu(x)=fgGm(r)$ versus the
dimensionless length $x = fgr$ for the two branches of the $B\sim
1$ (blue line), $B\sim 2$ (red line) and $B\sim 3$ (green line)
Skyrme Black Hole with $\alpha = 4\pi Gf^{2} = 0.001$ and
$x_{\triangle}=0.03$. The dotted lines are the unstable
solutions.}\label{muxh03}
\end{figure}
\begin{figure}[ht]
\epsfig{figure=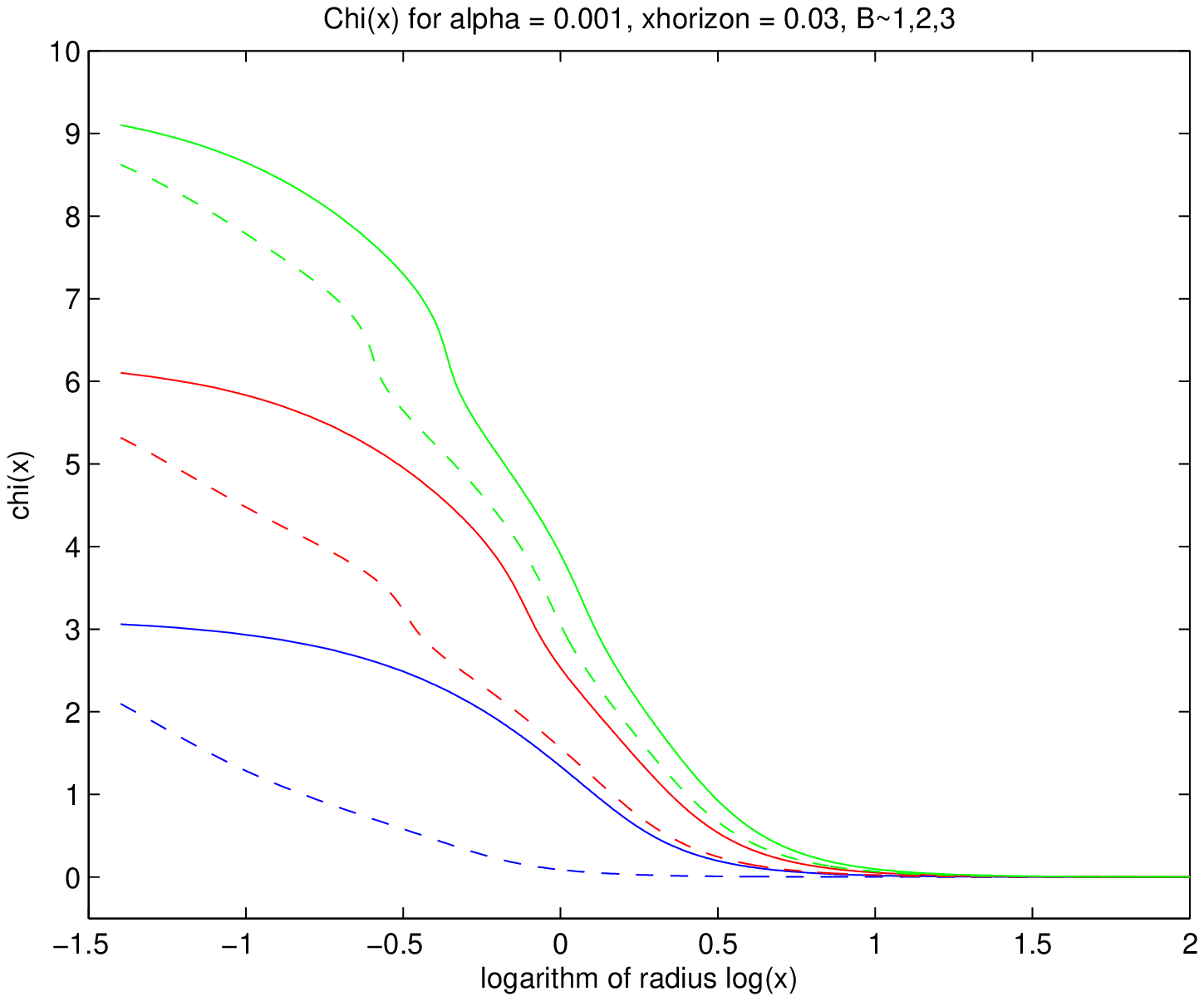,height=8cm,width=7cm} \caption{The
function $\chi(x)$ against dimensionless length $x=fgr$ for the
two branches of the $B\sim 1$ (blue line), $B\sim 2$ (red line)
and $B\sim 3$ (green line) Skyrme Black Hole with $\alpha = 4\pi
Gf^{2} = 0.001$ and $x_{\triangle}=0.03$. The dotted lines are the
unstable solutions.} \label{chixh03}
\end{figure}
The results of the numerical integration of the equations of
motion for the Einstein-Skyrme system are displayed in the figures
\ref{k001betas}-\ref{chixh03}.  From the figures it is easy to
read off the behaviour of the various functions appearing in the
conjectures.
\begin{enumerate}
    \item [{\bf{1.}}] $M_{\triangle}^{(n)}(r_{\triangle}) < M_{\triangle}^{(0)}(r_{\triangle})$ for all $n>0$ and all
    $r_{\triangle}$.
\end{enumerate}
Since the value of $M_{\triangle}^{(n)}(r_{\triangle})$ is just
half the value of the area under the curve $\beta(r_{\triangle})$
(Fig. \ref{k001betas}) and $M_{\triangle}^{(0)}$ is just given by
half the area under the curve $\beta = 1$ we can see that this
conjecture is true for all the values and solutions investigated.
Note that the validity of this conjecture just follows from the
validity of conjecture 2.
\begin{enumerate}
    \item [{\bf{2.}}] $\kappa^{(n)}(r_{\triangle}) < \kappa^{0}(r_{\triangle})$ for all $n>0$ and all
    $r_{\triangle}$.
\end{enumerate}
Using $\beta(r_{\triangle}) = 2r_{\triangle}\kappa(r_{\triangle})$
this can be seen to be true for all curves in Fig.
\ref{k001betas}. In addition since we have $\kappa =
\kappa_{s}A(1-2Gm')$ and $A(r)$ is always less than one (for
example see Fig. \ref{Axh03} and equation(\ref{Adash})) this
conjecture simply follows from the assumption of static, spherical
symmetry. It also implies conjecture 1 directly.
\begin{enumerate}
    \item [{\bf{3.}}] For a fixed value of $r_{\triangle}$ the horizon mass
    $M^{(n)}_{\triangle}$ and the surface gravity $\kappa^{(n)}$ are
    monotonically decreasing functions of $n$.
\end{enumerate}
This will hold for the stable solution branches but will not hold
for the unstable branches (see Fig. \ref{k001betas}), where both
$M^{(n)}_{\triangle}$ and $\kappa^{(n)}$ will be monotonically
increasing. One possible solution to this would be to argue that
the unstable branches should not be viewed as bound states of a
Schwarzschild black hole and the corresponding soliton but instead
should be viewed as bound states between the stable branch and the
corresponding unstable soliton. The binding energy would then be
\bea E_{bind} & = & M_{\triangle}^{unstab} - M_{\triangle}^{stab}
\nonumber \\ & & = \frac{1}{2G}\int^{r_{\triangle}}_{0}
(\beta^{unstab} - \beta^{stab})\d r_{\triangle}. \eea
As can be seen from Fig.\ref{k001betas}, the binding energy would
then decrease in magnitude as $n$ increases for a fixed horizon
size $r_{\triangle}$. However, for fixed $r_{\triangle}$ and
increasing $n$, while the mass of the unstable soliton would be
increasing, the mass of the stable black hole solution would be
decreasing. Since the mass of both objects in the bound system is
changing the original conjecture loses its motivation. It can also
be seen from Fig.\ref{k001betas} that for all cases considered the
magnitude of the binding energy will increase as $r_{\triangle}$
increases. Note also that in this case, if one still wished to
view the stable Skyrme black holes as bound systems between
Schwarzschild black holes and solitons, then one would effectively
be viewing the unstable Skyrme black holes as bound systems
between Schwarzschild black holes and both the stable and unstable
solitons.
\begin{enumerate}
    \item [{\bf{4.}}] $\beta ^{(n)}(r_{\triangle}) < 1$ for all $n>0$ and all
    $r_{\triangle}$.
\end{enumerate}
In the Einstein-Skyrme model the case $B=0$ is just the familiar
Schwarzschild solution. Thus $\beta^{(0)}=1$ and this is
essentially the same as conjecture 2 and Fig. \ref{k001betas}.
Notice also that since there is an upper bound on the radius of
the black hole $\beta^{(n)}(r_{\triangle})$ does not tend to one
asymptotically.
\begin{enumerate}
    \item [{\bf{5.}}] $M_{\triangle}^{(n)}$ is a monotonically increasing function
    of $r_{\triangle}$, is positive for all values of $n$ and
    vanishes at $r_{\triangle} = 0$.
\end{enumerate}
This will be true if $\mu'$ is never greater than a half and this
is certainly the case for all the solutions considered (see Fig.
\ref{muxh03}). For the case of the conjecture explicitly related
to the crossing point of the stable and unstable branches
\beq \label{closedloop} M_{sol}^{unstab}-M_{sol}^{stab}  =
\frac{1}{2G}\oint\beta(r_{\triangle})\d r_{\triangle}\eeq
the following values can be computed using the numerical
solutions.\bigskip

\begin{tabular}{|c|c|c|c|c|}
  \hline
  $B$ & $M^{stab}_{sol}$ & $M^{unstab}_{sol}$ & $\frac{1}{2G}\oint \beta(r)\d r$ & Difference \\
  \hline
  $B\sim 1$ & $0.00578$
 & 0.02637
 & 0.02078
 & 0.00019
 \\
  $B\sim 2$ & 0.01706
 & 0.03223
 & 0.01532
 & 0.00016

 \\
  $B\sim 3$ & 0.03363
 & 0.04331
 & 0.00979
 & 0.00011
 \\
  \hline
\end{tabular}
\bigskip

While there is a slight mismatch in the left-hand side and
right-hand side it is most likely that this is simply due to the
numerical accuracy of the numerical solutions. The convergence of
the numerical integration on the right hand side of
(\ref{crossint}) was checked and was found to converge quickly to
the values given in the table.

\section{Conclusions}

We have investigated the static black hole conjectures stemming
from the isolated horizons formalism using numerically generated
solutions to the Einstein-Skyrme model. We have found that the
numerical results are in impressive agreement with the
conjectures. The only mismatch is in the behaviour of
$\beta(r_{\triangle})$ for the linearly unstable branches.
However, since conjecture three is the only conjecture that relies
critically on the assumption of the hairy black holes being bound
states between bare black holes and solitons, it would be useful
to investigate whether this behaviour is repeated in other models
such as the Einstein-Proca system. It is possible that this
conjecture will fail generically for models exhibiting the
two-branch-merging behaviour seen in the Skyrme model and will
need to be replaced in these cases with something else, such as
the modification discussed here, whereby the unstable branches are
viewed as bound states of the solitons and the stable black holes.
Further investigations would however need to be carried out to
test this idea. While we have restricted ourselves to spherically
symmetric, asymptotically flat solutions, similar tests could be
carried out on Skyrme black holes with negative cosmological
constant \cite{Shiiki:2005aq} provided that due care is taken that
the Hamiltonian formalism is well defined in the non-flat
asymptotic region. However, it is not clear how to extend the
analysis to the axisymmetric $B\sim 2$ solutions of
\cite{Sawado:2004yq} since these have a lower bound on the horizon
radius and hence are not connected to their corresponding regular
soliton solutions and it is not clear in this situation how to
define the horizon mass. Since these axisymmetric solutions
represent minimal energy solutions in the $B\sim 2$ sector there
may be some interesting relation to the $BPS$ bound. In addition,
it would be interesting to see whether anything could be said
about the lowest energy configurations in the $B>2$ sector since
these are expected to have discrete tetrahedral and octahedral
symmetries.

\section{Acknowlegdements}
We would like to thank Marcelo Salgado and Daniel Sudarsky for
helpful comments and discussions. This work was supported by the
Marsden Fund of the Royal Society of New Zealand.


\begin{thebibliography}{999}

\bibitem{Ashtekar:2000sz}
  A.~Ashtekar, C.~Beetle, O.~Dreyer, S.~Fairhurst, B.~Krishnan, J.~Lewandowski and J.~Wisniewski,
  Phys.\ Rev.\ Lett.\  {\bf 85} (2000) 3564
  [arXiv:gr-qc/0006006].

\bibitem{Ashtekar:2000nx}
  A.~Ashtekar, A.~Corichi and D.~Sudarsky,
  Class.\ Quant.\ Grav.\  {\bf 18} (2001) 919
  [arXiv:gr-qc/0011081].

\bibitem{Corichi:2000dm}
  A.~Corichi, U.~Nucamendi and D.~Sudarsky,
  Phys.\ Rev.\ D {\bf 62} (2000) 044046
  [arXiv:gr-qc/0002078].

\bibitem{Luckock:1986em}
  H.~Luckock,
  in ``\emph{String theory, quantum cosmology
  and quantum gravity, integrable and conformal invariant
  theories: Proceedings of the Paris-Meudon Colloquim, 22-26
  September 1986.}" (World Scientific, Singapore, 1987)

\bibitem{Bizon:1992gb}
  P.~Bizon and T.~Chmaj,
  Phys.\ Lett.\ B {\bf 297} (1992) 55.

\bibitem{Droz:1991cx}
  S.~Droz, M.~Heusler and N.~Straumann,
  Phys.\ Lett.\ B {\bf 268} (1991) 371.

\bibitem{Volkov:1998cc}
  M.~S.~Volkov and D.~V.~Gal'tsov,
  Phys.\ Rept.\  {\bf 319} (1999) 1
  [arXiv:hep-th/9810070].

\bibitem{Breton:2003tk}
  N.~Breton,
  Phys.\ Rev.\ D {\bf 67} (2003) 124004
  [arXiv:hep-th/0301254].

\bibitem{Corichi:2005pa}
  A.~Corichi, U.~Nucamendi and M.~Salgado,
  Phys.\ Rev.\ D {\bf 73} (2006) 084002
  [arXiv:gr-qc/0504126].

\bibitem{Kleihaus:2000kv}
  B.~Kleihaus and J.~Kunz,
  Phys.\ Lett.\ B {\bf 494} (2000) 130
  [arXiv:hep-th/0008034].

\bibitem{Ibadov:2005rb}
  R.~Ibadov, B.~Kleihaus, J.~Kunz and M.~Wirschins,
  Phys.\ Lett.\ B {\bf 627} (2005) 180
  [arXiv:gr-qc/0507110].

\bibitem{Sawado:2004yq}
  N.~Sawado, N.~Shiiki, K.~i.~Maeda and T.~Torii,
  Gen.\ Rel.\ Grav.\  {\bf 36} (2004) 1361
  [arXiv:gr-qc/0401020].

\bibitem{Heusler:1992av}
  M.~Heusler, S.~Droz and N.~Straumann,
  Phys.\ Lett.\ B {\bf 285} (1992) 21.

\bibitem{Heusler:1993ci}
  M.~Heusler, N.~Straumann and Z.~H.~Zhou,
  Helv.\ Phys.\ Acta {\bf 66} (1993) 614.

\bibitem{Wald:book}
  R.~M.~Wald, \emph{General Relativity}, (The University of Chicago
  Press, Chicago, 1984)

\bibitem{Hartmann:2001ic}
  B.~Hartmann, B.~Kleihaus and J.~Kunz,
  Phys.\ Rev.\ D {\bf 65} (2002) 024027
  [arXiv:hep-th/0108129].

\bibitem{Shiiki:2005aq}
  N.~Shiiki and N.~Sawado,
  Phys.\ Rev.\ D {\bf 71} (2005) 104031
  [arXiv:gr-qc/0502107].

\end{thebibliography}
\end{document}